\documentclass[pdflatex, sn-vancouver-num]{sn-jnl}

\usepackage{graphicx}
\usepackage{multirow}
\usepackage{amsmath, amssymb, amsfonts}
\usepackage{amsthm}
\usepackage{mathrsfs}
\usepackage[title]{appendix}
\usepackage{xcolor}
\usepackage{textcomp}
\usepackage{manyfoot}
\usepackage{booktabs}
\usepackage{algorithm}
\usepackage{algorithmicx}
\usepackage{algpseudocode}
\usepackage{listings}

\usepackage{dsfont}
\usepackage{dcolumn}
\usepackage{braket}
\usepackage{bm}
\usepackage{tikz}
\usepackage{hyperref}
\usepackage[nameinlink, capitalize]{cleveref}

\hypersetup{%
	plainpages=false,
	colorlinks=true,
	linkcolor=[RGB]{0, 114, 195},
	citecolor=[RGB]{0, 114, 195},
	urlcolor=[RGB]{0, 114, 195},
	pdfborder={0 0 0},
	breaklinks=true,
	bookmarksnumbered=true,
	bookmarksopen=true,
}

\creflabelformat{equation}{#2#1#3}

\theoremstyle{definition}
\newtheorem*{definition*}{Definition}

\begin{document}
    \title[Article Title]{%
        Reinforcement learning entangling operations on spin qubits
    }
    \author[1, 2]{\fnm{Mohammad}\sur{Abedi}}\email{m.abedi@fz-juelich.de}
    \author[1, 3]{\fnm{Markus}\sur{Schmitt}}\email{mar.schmitt@fz-juelich.de}
    \affil[1]{
        PGI-8 (Quantum Control),
        Forschungszentrum J\"ulich,
        Wilhelm-Johnen-Straße,
        52428,
        J\"ulich
    }
    \affil[2]{%
        Fakult\"at f\"ur Physik,
        University of Regensburg,
        Universit\"atsstraße 31,
        D-93051,
        Regensburg
    }
    \affil[3]{%
        Fakult\"at f\"ur Informatik und Data Science,
        University of Regensburg,
        Universit\"atsstraße 31,
        D-93040,
        Regensburg
    }
    \date{\today}

    \abstract{%
        High-fidelity control of one- and two-qubit gates past the error
        correction threshold is an essential ingredient for scalable quantum
        computing. We present a reinforcement learning (RL) approach to find
        entangling protocols for semiconductor-based singlet-triplet qubits in a
        double quantum dot. Despite the presence of realistically modelled
        experimental constraints, such as various noise contributions and finite
        rise-time effects, we demonstrate that an RL agent can yield
        performative protocols, while avoiding the model-biases of traditional
        gradient-based methods. We optimise our RL approach for different
        regimes and tasks, including training from simulated process tomography
        reconstruction of unitary gates, and investigate the nuances of RL agent
        design.
    }

    \maketitle

    \section{Introduction}
    Quantum technology is a rapidly growing research field with extensive
    cross-disciplinary interests. Its applications for computation is
    conditioned on the implementation of accurate logic gates. Two facets of
    this requirement are the development of quantum hardware with fast, scalable
    coherent dynamics, and the ability to steer and manipulate the system to
    implement arbitrary logical operations, even in the presence of decohering
    noise. The theory of quantum control (QC) and quantum optimal control
    (QOC)~\cite{butkovskii1979control, 1979control, glaser2015training} has
    provided the theoretical framework for the latter task. A typical approach
    to (optimal) quantum control usually involves optimisation through the
    computation of gradients of a Hamiltonian~\cite{khaneja2005optimal,
    de2011second, krotov1995global, machnes2015gradient}. Although this approach
    is efficient, it's stipulated on the accuracy of the model (i.e. its
    Hamiltonian description) and, in the case of real quantum environments, on
    the validity of a noise description. Such model bias may cause
    gradient-based optimisation to perform sub-optimally or converge to a local
    minimum.

    In recent years, reinforcement learning (RL)~\cite{sutton2018reinforcement}
    has swiftly developed into a general framework for (optimal) control and
    interactive learning tasks with a diverse domain of utilisation such as
    autonomous driving~\cite{kiran2021deep}, robotics~\cite{smith2022walk},
    fine-tuning large-language-models with human
    feedback~\cite{ouyang2022training}, and learning to play games such as Chess
    and Go~\cite{Schrittwieser_2020}. This set of algorithms is deployed to
    optimise sequential decision-making and maximise cumulative feedback. No
    prior knowledge of the underlying system is required, since learning and
    experience are achieved simultaneously through an interactive process
    between an agent and its environment. These model-free (i.e.
    Hamiltonian-free) approaches have already successfully been implemented in
    the context of QC~\cite{tsubouchi2008rovibrational, 8759071,
    PRXQuantum.2.040324, niu2019universal, nguyen2024reinforcement, 10040740,
    bukov2018reinforcement, an2019deep, sivak2022model}. However, there are a
    number of remaining important challenges that we investigate.

    \begin{figure}[h]
        \centering
        \includegraphics[scale=0.73]{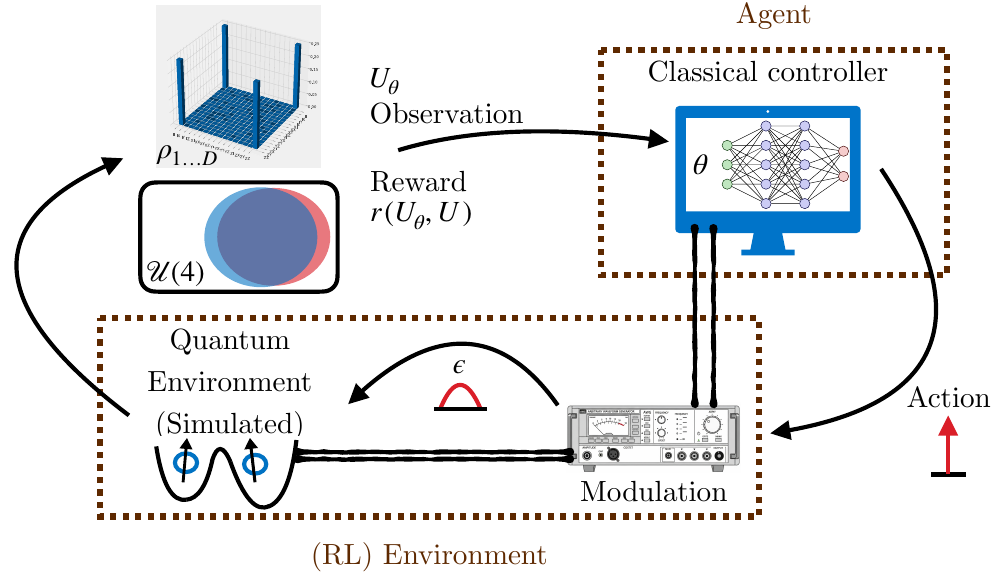}
        \caption{%
            The interaction loop of reinforcement learning (RL) applied to
            quantum control (QC). An RL agent (i.e. a neural network in a
            classical computer) outputs a continuous number within a bounded
            interval (i.e. an action) in instant time. Due to finite-rise
            effects and ringing of controlled hardware, the action is convolved
            with a measured impulse response function to better represent the
            model dynamics. The double quantum dot system is then perturbed
            using a Trotterised decomposition scheme. Feedback is provided to
            the agent in the form of an observation characterizing an implicit
            action-evolution correlation, and a reward representing an
            observation-action dependent goal. An observation consists of either
            complete or limited (e.g. process tomography) information about the
            current generated unitary, while a reward is a function of the
            current unitary and a target gate. Finally, the agent outputs a new
            feedback-dependent action.
        }\label{fig:rl_qc}
    \end{figure}

    In contrast to classical control systems, the observation of a state in a
    quantum environment presents a fundamental limitation in the ability of
    environment-agent interactions since any (weak or otherwise) measurement
    perturbs the system. Alternatively, preparing many identical copies of a
    known initial state, followed by implementing a control protocol, and then
    conducting a complete measurement requires substantial experimental
    resources. This is because process tomography typically requires millions of
    measurements, even for a small qubit system~\cite{torlai2023quantum,
    PRXQuantum.2.040324}. Further complications are certain experimental
    features in different quantum platforms such as delayed response to applied
    control, which, from the RL perspective, affects the Markovian properties of
    the problem. Additionally, decohering noise including time-correlated noise
    compounds with the statistical noise of measurement, adding a significant
    amount of uncertainty in the optimisation process.

    Previous work in QC with RL included discrete-pulse optimisation of the IBM
    superconducting platform, finding more time-efficient gates with
    RL~\cite{PRXQuantum.2.040324}, synthesising noise-resilient pulses in gmon
    systems given knowledge of a noisy environment~\cite{niu2019universal},
    constructing entangling gates on simulated transmon systems with continuous
    controls~\cite{nguyen2024reinforcement}, and demonstrating agent-based
    learning ability to discover novel protocols compared to conventional
    ans\"atze. Other work includes synthesising unitaries with discrete-pulse
    sequences~\cite{an2019deep, 10040740}, investigating optimal control
    protocols in different controllability
    regimes~\cite{bukov2018reinforcement}, and optimising control sequences of
    parameterised circuits to generate desired gates with measurement
    feedback~\cite{sivak2022model}. Another investigation used RL ans\"atze with
    a Hamiltonian-gradient-based approach to yield improved
    fidelities~\cite{sarma2025designing}. Our work is in the spirit of these
    efforts and our aim is to further bridge the gap between comparatively
    abstract control models and a viable quantum platform that contains the
    diverse nuances expected of an experimental device.

    As a concrete example, we will consider semiconductor-based spin
    qubits~\cite{RevModPhys.95.025003} and in particular singlet-triplet
    qubits~\cite{Cerfontaine:768380, Cerfontaine_2020}. Platforms based on spin
    qubits are a suitable candidate for quantum computing as they have long
    coherence times~\cite{muhonen2014storing, PhysRevLett.110.146804},
    experimentally demonstrated gate fidelities past the error correction
    threshold for both one and two qubit gates~\cite{yoneda2018quantum,
    noiri2022fast}, allow for fast electrical control~\cite{Cerfontaine_2020},
    and are scalable due to existing knowledge in industrial
    fabrication~\cite{PRXQuantum.4.020305}. A sophisticated gradient-based
    approach to design optimal entangling operations using a detailed model
    including the relevant device imperfections has been developed previously by
    Cerfontaine~\textit{et al.}~\cite{Cerfontaine_2020}, providing a competitive
    benchmark for our RL approach. Furthermore, although not the focus of our
    investigation, closed-loop control with fine-tuning with experimental data
    has been performed for GaAs qubits~\cite{cerfontaine2020closed}. Future work
    could incorporate some or all parts of this workflow with RL agents.

    We examine the design choices of RL agents and their impact on performance,
    emphasising moving towards a practical approach to quantum control. These
    choices are informed by the experimental features of finite-rise times,
    multiple sources of decohering noise, and shot noise. Such design choices
    include exploring model- and history-of-action-based belief states to learn
    noise-robust entangling strategies. We aim to find solutions that exceed the
    accuracy of the gradient-based approach, sequentially surveying the
    experimental aspects of a quantum dot platform, with increasing complexity.

    In the following, we will first introduce the details of the physical
    platform in~\cref{sec:qd} and basics of RL in~\cref{sec:rl}, before
    presenting the results in~\cref{sec:results}.

    \section{Quantum Dots}\label{sec:qd}
    We present a brief overview of the quantum platform that is optimised.
    Readers interested in the full experimental details are referred
    to~\cite{Cerfontaine:768380, Cerfontaine_2020}. One- and two-qubit gates
    were previously synthesised using the filter function
    formalism~\cite{Hangleiter_2021} and the gradient-based Levenberg–Marquardt
    algorithm to optimise pulses which minimise the average gate infidelity. Our
    interest in testing optimal controllability of this particular model with an
    RL agent is motivated by experimental constraints of noise and finite
    rise-time effects which provide additional challenges in the learning
    process, and the availability of a comparative
    benchmark~\cite{Cerfontaine_2020}.

\begin{figure}
        \centering
        \includegraphics[scale=1.0]{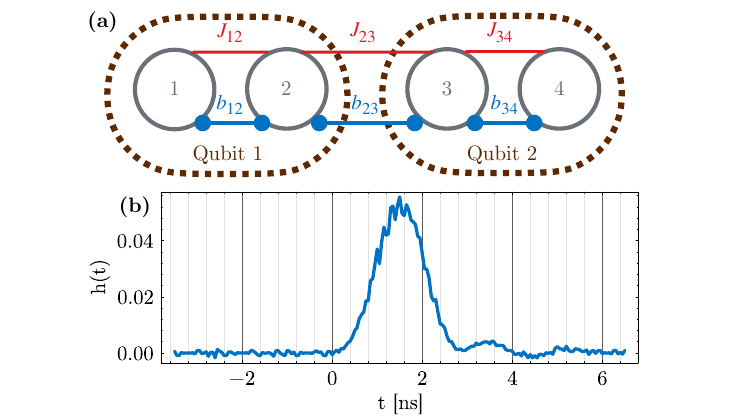}
        \caption{%
            (a) A two-qubit system using four adjacent electron spins in a
            semiconducting heterostructure. Voltage pulses between dots are
            controlled to modify the strength of the exchange interactions
            \(J_{ij}\), affecting the occupation levels of computational
            subspace.\@ \(b_{ij}  = B_{i} - B_{j}\) are the average magnetic
            field gradients across dots \((i, j)\), and
            \(\frac{1}{4}\sum_{i}B_{i} = B_{G}\) gives the total applied
            external magnetic field. (b) The kernel used to convolve
            piecewise-constant pulses creating smoothened pulses which more
            accurately describe the qubit dynamics. Notably, there is a
            \(\sim\hspace{-0.3em}1\text{ns}\) delay from the input by the
            controller. This is relevant when the inputs are sent in less than
            this time, as it can cause symbol interference.
        }\label{fig:qd_impulse}
    \end{figure}

    This singlet-triplet qubit system consists of four adjacent, confined
    quantum dots in a semiconducting GaAs heterostructure (\cref{fig:qd_impulse}
    (a)) which are used to encode a (logical) two-qubit system. The system is
    described by the following Hamiltonian, given in units of angular
    frequencies (\(\hbar = 1\)):
    \begin{equation}\label{eq:hamiltonian_qd}
        \begin{split}
            H = & {%
                \sum^{3}_{i}{%
                    \frac{J(\epsilon_{i, i + 1})}{4}
                    \overline{\sigma}^{(i)}
                    \cdot
                    \overline{\sigma}^{(i + 1)}
                }
            }\\
            &+ \frac{B_{G}}{2}\sum^{4}_{i}\sigma^{(i)}_{z}\\
            &+ {%
                \frac{b_{12}}{8}
                \left(
                    -3\sigma^{(1)}_{z}
                    + \sigma^{(2)}_{z}
                    + \sigma^{(3)}_{z}
                    + \sigma^{(4)}_{z}
                \right)
            }\\
            &+ {%
                \frac{b_{23}}{4}
                \left(
                    -\sigma^{(1)}_{z}
                    - \sigma^{(2)}_{z}
                    + \sigma^{(3)}_{z}
                    + \sigma^{(4)}_{z}
                \right)
            }\\
            &+ {%
                \frac{b_{34}}{8}
                \left(
                    -\sigma^{(1)}_{z}
                    - \sigma^{(2)}_{z}
                    - \sigma^{(3)}_{z}
                    + 3\sigma^{(4)}_{z}
                \right)
            }\\
        \end{split}
    \end{equation}
    Here, \(\sigma_{a}^{(i)}\) is the Pauli operator that acts on the quantum
    dot with label \(i\). Each spin experiences a constant magnetic field
    \(B_{i}\). \(B_{G} = \sum_{i}B_{i}\) is the mean field and \(b_{ij} = B_{i}
    - B_{j}\) represents the magnetic gradient across adjacent dots. Application
    of an external magnetic field energetically splits the six states with
    \(m_{s} = 0\) and those with non-zero magnetic moment. These six states are
    further split into a computational subspace:
    \begin{equation}
        \left \{
            \begin{aligned}
                \ket{00} &:= \ket{\uparrow\downarrow\uparrow\downarrow}\ \
                \ket{01} := \ket{\uparrow\downarrow\downarrow\uparrow}\\
                \ket{10} &:= \ket{\downarrow\uparrow\uparrow\downarrow}\ \
                \ket{11} := \ket{\downarrow\uparrow\downarrow\uparrow}
            \end{aligned}
        \right \}
    \end{equation}
    Additionally, there are two inter-double dot leakage states accessible via
    \(J_{23}\):
    \begin{equation}
        \left \{
            \ket{L_{1}} := \ket{\uparrow\uparrow\downarrow\downarrow}\ \
            \ket{L_{2}} := \ket{\downarrow\downarrow\uparrow\uparrow}\\
        \right \}
    \end{equation}
    Detuning voltage pulses \(\epsilon_{ij}\) are used to affect the energy
    difference between adjacent dots, controlling the exchange interaction
    \(J_{ij}\), which is modelled as \(J_{ij} =
    J_{0}\exp{(\frac{\epsilon_{ij}}{\epsilon_{0}})}\), where \(J_{0},
    \epsilon_{0}\) are the base units of this system, given in ns and meV
    respectively.

    \(\epsilon_{ij}(t)\) is controlled by an arbitrary waveform generator (AWG)
    which is constrained by amplitude bounds as well as a fixed sample period
    \(T_{s}\), producing piecewise-constant pulses. Furthermore, the finite
    bandwidth of the system combined with ringing of the control hardware is
    modelled by the use of a convolution filter smoothening piecewise-constant
    pulses
    \begin{equation}\label{eq:convolution}
        \epsilon'_{ij}(t) = \int^{t}_{0}\epsilon_{ij}(t)h(t - t')\text{d}t'
    \end{equation}
    where \(h(t)\) is the convolution kernel. We use the measured impulse
    response function (\cref{fig:qd_impulse} (b)) as our convolution kernel.
    This was obtained by digitising data from~Cerfontaine~\textit{et
    al.}~\cite{Cerfontaine_2020}, which was constructed with the following
    procedure: A square pulse was sent to the device as an input via an AWG
    through a series of low-loss cables. This yielded a measured distorted
    square pulse, and the time-derivative of the resulting step response
    corresponds to the impulse response function\footnote{As an alternative, the
    impulse response function can be measured directly using instant
    (Dirac-delta) pulses, whose integral corresponds to the step response.}.
    This impulse response function is used to convolute the input pulses in our
    model.

    Three noise contributions to the system are included in the optimisation
    procedure. One of these is the quasi-static hyperfine noise on drift mean
    magnetic field gradients \(b_{ij}\). This noise fluctuates on time scales
    slower than the protocol duration. Hence, it is modelled as a Gaussian noise
    with standard deviation \(\sigma_{b}\) for each individual protocol:
    \begin{equation}
        \tilde{b}_{ij} = {%
            b_{ij}
            + \delta_{b_{ij}};
            \ \ \ \delta_{b_{ij}}\sim \mathcal{N}(0, \sigma_{b})
        }
    \end{equation}
    A similar slow charge noise with standard deviation \(\sigma_{\epsilon}\)
    contributes to one of two sources of noise on the detuning voltages
    \(\epsilon_{ij}\). Furthermore, the decoherence noise from \(\epsilon_{ij}\)
    is also affected by fast (nanosecond scale) charge noise, and is modelled
    via a full noise spectrum of time-correlated coloured noise
    (\(\sim\frac{1}{f^{\alpha}}\) for frequency \(f\)). An optimistic scenario
    with minimal effects from high frequency regimes can be represented by
    taking \(\alpha = 0.7\) and a pessimistic extreme is represented as \(\alpha
    = 0.0\), or white noise:
    \begin{equation}\label{eq:noise}
        \begin{split}
            \tilde{\epsilon}_{ij}(t) = {%
                \epsilon^{\prime}_{ij}(t)
                + \delta^{1}_{\epsilon^{\prime}_{ij}}
                + \delta^{2}_{\epsilon^{\prime}_{ij}}(t);
            }\\
            \ \ \ \delta^{1}_{\epsilon_{ij}}
            \sim\mathcal{N}(0,\ \sigma_{\epsilon})
            \ \ \ \delta^{2}_{\epsilon^{\prime}_{ij}}
            \sim S_{\epsilon, \alpha}(f)
        \end{split}
    \end{equation}

    System parameters are given in terms of \(\epsilon_{0}\) for detuning
    voltages and \(J_{0}\) for magnetic field and exchange interactions. For a
    GaAs device, we use the previously used parameters~\cite{Cerfontaine_2020}
    and set \(J_{0} = 1\text{ns}^{-1}\) with \(b_{12} = -b_{34} = J_{0}\) and
    \(b_{23} = 7J_{0}\) to suppress leakage by making the exchange interaction
    between the qubits costly~\cite{PhysRevB.90.045418}. Slow fluctuations on
    the magnetic field gradients are \(\sigma_{b} = 0.0105J_{0}\). The detuning
    pulses are constrained by the AWG with a rise time of
    \(\sim\hspace{-0.3em}1\text{ns}\) and are hence bounded to the interval
    \([-5.4\epsilon_{0}, 2.4\epsilon_{0}]\) where \(\epsilon_{0} =
    0.272\text{mV}\) and include a slow charge contribution of standard
    deviation \(\sigma_{\epsilon} = 0.0294\epsilon_{0}\). The fast charge noise
    spectrum is extrapolated to \(4\times 10^{-20}\text{V}^{2}\text{Hz}^{-1}\)
    at 1 MHz such that \(S_{\epsilon, \alpha}(f) = 53.8\epsilon_{0}^{2}\text{ns}
    {\left(\frac{1\text{Hz}}{f}\right)}^{\alpha}\).

    For optimisation, we train an agent with a numerically simulated quantum dot
    environment. We aim to find control pulses that construct an entangling
    gate, CNOT, in the logical basis with a synthesised and discretised sequence
    of detuning pulses \({\{\epsilon_{ij, t}\}}^{t=N}_{t = 0}\) generated in
    some finite time \(T\) consisting of \(N\) segments. This gives a sample
    period for the AWG as \(T_{s} = \frac{T}{N}\).

    In our numerical simulation of the device dynamics, the unitary is generated
    using the Trotterised evolution of the
    Hamiltonian~\eqref{eq:hamiltonian_qd}. The accuracy of the discretised time
    evolution is given by the time step, \(\mathcal{O}(\delta t^{2})\), and is
    chosen to be sufficiently small to mitigate discretisation errors. By
    oversampling the pulses \(n\) times (i.e. for each input, there are \(n\)
    substeps), we produce increasingly smoother convoluted traces of the input
    pulses~\eqref{eq:convolution} and fix the time step as \(\delta t = T_{s} /
    n = T / Nn\) with the approximate time evolution operator \(U(t) =
    \prod^{M}_{m = 0} \exp{\left(-i\delta tH(\epsilon'_{ij, m})\right)}\) for
    \(M = Nn\) steps and shaped (i.e. convoluted) pulses \(\epsilon'_{ij, m}\).
    For each step, we have \(U_{m} = \exp{\left(-i\delta t
    H(\epsilon^{\prime}_{ij, m})\right)}U_{m - 1} \) and \(U_{0} = \mathds{1}\).
    Note to distinguish the input sequence discretisation, which is labelled by
    \(t = 1\ldots N\) (i.e. the input unshaped pulse amplitudes), from the
    numerical simulation (including oversampling), which is labelled by \(m =
    1\ldots Nn\).

    Furthermore, we fix the final four segments at their minimum value to ensure
    each protocol ends with minimal exchange interaction. This is an enforced
    constraint on the optimisation procedure, giving us \(N - 4\) choices of
    pulse amplitudes.

    Taking into account the effects of finite rise time, noise contributions,
    and model simulation, we can now proceed to introduce the basic concepts of
    the RL optimisation algorithm.

    \section{Reinforcement Learning on Quantum Dots}\label{sec:rl}
    \begin{table}[t]
        \centering
        \begin{tabular}{|c|c|}  
            \hline  
            \textbf{RL} & \textbf{Quantum Control} \\
            \hline  
            Observation \(o_{t}\) & Pulse history \(\epsilon_{o:t}\) and / or
            unitary gate \(U_{t}\) \\
            & (potentially estimated with process \\ & tomography
            statistics), and time-to-go \\
            & \\
            Belief \(b_{t}\) & A belief state over the underlying latent \\
            & unitary conditioned on the history of \\
            & actions and observations \\
            & \\
            Action \(a_{t}\) & A real continuous value in a closed interval \\
            & that is smeared via an AWG to become a \\
            & pulse amplitude \(\epsilon \) \\
            & \\
            Reward \(r_{t}\) & A function of the (estimated) gate fidelity \\
            & (\cref{eq:nlif})\\
            \hline  
        \end{tabular}
        \caption{%
            RL components and their quantum control aspects.
        }\label{tab:rl_qc}
    \end{table}
    We present a brief description of reinforcement learning (RL) and its
    formulation in the context of quantum dots. A more comprehensive overview is
    given in~\hyperref[ssec:rl_appendix]{Appendix A}, and Sutton \&
    Barto~\cite{sutton2018reinforcement}.

    In RL, an abstract agent or learning system receives feedback signals from
    its interactions with an environment, learning an optimal strategy or
    sequence of interactions to achieve a desired task. RL does not require
    expert supervision or knowledge distillation; an agent can learn directly
    from experience, and only needs to learn to maximise the feedback signal. A
    model of the environment is not explicitly required.

    In the framework of \textit{Markov decision processes
    (MDPs)}~\cite{sutton2018reinforcement} underlying RL, an agent learns and
    interacts with an environment by taking actions \(a_{t}\in\mathcal{A}\)
    through a conditional probability distribution known as a \textit{policy}
    \(\pi(a_{t}|s_{t})\), where \(s_{t}\in\mathcal{S}\) is a state of the
    environment. This action affects the environment state through a Markovian
    transition probability function \(\mathcal{T}(s_{t + 1}|s_{t}, a_{t})\),
    giving a new state \(s_{t + 1}\). The agent then observes this updated state
    alongside a reward signal \(r_{t}\) from the environment and its goal is to
    learn a policy \(\pi^{*}\) that maximises the cumulative reward over a
    sequence of actions, the \textit{return}. A sequence may be constructed to
    terminate upon reaching a particular state or after a fixed number of steps.
    A finite sequence is known as an \textit{episode} and an agent may require
    many episodes to successfully learn an optimal strategy.

    In quantum systems, direct access to the environment is limited due to the
    distinctive nature of measurement. Therefore we generalise the MDP framework
    to a \textit{partially observable Markov decision process
    (POMDPs)}~\cite{sutton2018reinforcement}. In addition to the set of states
    \(\mathcal{S}\), it includes an observation set \(\mathcal{O}\) and an
    observation probability function of getting an observation \(o_{t}\) given a
    state-action pair \(\mathcal{Z}(o_{t + 1}|s_{t + 1}, a_{t})\). This added
    stochasticity adds uncertainty to the agent's policy as a state model is
    additionally required. One choice of a state model is to maintain a
    \textit{history} of actions and observations \(h_{t} =
    a_{0}o_{1}a_{1}o_{2}\ldots a_{t - 1}o_{t}\) which satisfies the Markovian
    property for being an RL state. However, due to potential intractability of
    histories for long sequences, a compact representation of the history known
    as a \textit{belief state} \(b_{t} = f^{\star}(h_{t})\) is sufficient
    (albeit optimisation may be difficult). A belief state is then a probability
    distribution over the underlying states \(s_{t}\) conditioned on the history
    \(h_{t}\).

    The engineering of an RL loop is a non-trivial task and requires careful
    consideration of the state, action, and reward spaces to solve a desired
    task. We present an overview of our design choices for each category below
    and in~\cref{sec:results}. A summary is given in~\cref{tab:rl_qc} and an
    overview can be seen in~\cref{fig:rl_qc}. A (classical) controller is
    constituted by a parameterised neural network policy \(\pi_{\theta}\) that
    outputs a continuous action in a closed interval between a chosen minimal
    and maximal pulse amplitude, conditioned on the obtained observation. This
    action is then passed to the quantum system through an AWG device, which
    constitutes a part of the RL environment dynamics. Finally, the system
    evolves under the simulated dynamics with Hamiltonian
    (\cref{eq:hamiltonian_qd}) and noise (\cref{eq:noise}). This loop is
    repeated for a fixed number of time steps and we construct a protocol
    (corresponding to an episode).

    For a concrete realisation of our RL approach, we employ the widely used
    Soft-actor-critic (SAC) algorithm~\cite{haarnoja2018soft, haarnoja2019soft},
    an entropic-exploration based approach which has demonstrated good
    performance for model-free continuous control, combined with dropout
    Q-networks~\cite{hiraoka2022dropoutqfunctionsdoublyefficient} to improve
    sample efficiency, and distributional value estimators to reduce
    overestimation bias~\cite{kuznetsov2020controlling}. A brief overview of
    this policy gradient actor-critic model can be found
    in~\hyperref[ssec:rl_appendix]{Appendix A}, and a list of hyperparameters
    in~\hyperref[ssec:rl_hp]{Appendix B}.
    \subsection{States and observations}
        For the purpose of quantum gate control, the agent's observation of the
        environment requires information about the unitary operator. We
        investigate the extent of information about the unitary \(U_{t}\) and
        its effect on agent performance and interpolate between providing
        minimal information and performing an exact tomographic reconstruction.

        One choice is an observation that is constructed from the history of
        inputted pulses (i.e.\@ \(o_{t} = \epsilon_{0:t}\)) and represents a
        minimum limit of information acquired about the unitary. Since there is
        a non-injective mapping from pulses to a unitary operator
        \(\epsilon_{0:t} \mapsto U_{t} = U(\epsilon_{t - 1})U(\epsilon_{t -
        2})\ldots \mathds{1}\) (this is similar to a belief state
        representation), we hypothesise (and test) that the agent may still be
        able to learn using this information. In this opposing limit, a choice
        of directly inputting the unitary (i.e.\@ \(o_{t} = U_{t}\)) into the
        observation represents maximal informational acquisition. This
        corresponds to an exact tomography of a unitary. In the intermediate
        regime, a choice of performing a (simulated) quantum process tomography
        and collecting measurement statistics yields a noisy observation and
        partial information about the unitary. For this latter case, we use a
        minimal set of positive operator-valued measures (POVMs) to reconstruct
        the unitary operator (\hyperref[ssec:u_t]{Appendix C}).

        Additionally, since an agent has no apriori knowledge if the number of
        time steps and the total protocol time are fixed, we must include the
        time-to-go~\cite{pardo2022timelimitsreinforcementlearning} in the RL
        observation, as it is a time-limited task.
    \subsection{Actions}
        The action space is a continuous set of real values in a closed interval
        \(\mathcal{A} = [\epsilon_{\text{min}}, \epsilon_{\text{max}}]\). The
        choice of a continuous action space is motivated by the desire for a
        fine-grained control of the pulse amplitudes. However, alternative
        designs may impose further constraints. For example, the action space
        could be a closed interval \([-\Delta\epsilon, \Delta\epsilon]\) that
        changes the current pulse amplitude by the chosen value, restricting the
        protocol space to smoothly varying
        pulses~\cite{nguyen2024reinforcement}. For simplicity, we only use the
        simplest case of a continuous bounded value.
    \subsection{Rewards}
        As a choice of a reward signal or figure of merit, we use a function of
        the negative logarithmic gate infidelity (NLIF), a simple distance
        measure between our generated gate and the desired gate (e.g. CNOT):
        \begin{equation}\label{eq:nlif}
            f = -\log_{10}\left(
                1
                - \left|
                    \frac{%
                        \text{Tr}\left(
                            U_{\text{target}}^{\dagger}U_{\text{final}}
                        \right)
                    }{d}
                \right|^{2}
            \right)
        \end{equation}
        Here, a logarithmic scale is used ensuring that the agent receives a
        distinguishable signal even at high fidelities.

        Despite its simplicity, the presence of decohering and shot noise, both
        of which may be present when obtaining the final gate, presents a
        challenge to reward functions based on the NLIF. The fidelity scales
        with the system noises up to a quartic term~\cite{Cerfontaine_2020}
        while scaling exponentially with number of shots with the
        measurement noise (\hyperref[ssec:u_t]{Appendix C}).

        We aim to identify protocols that are robust with respect to these noise
        sources. One approach is to constantly monitor the effects of noise on
        gate generation and adjust the control pulses accordingly. For this
        purpose, we can use a reward function of the NLIF given at the final
        time-step. However, more realistically, we intend to find control pulses
        that perform well over many noise realisations, and therefore we use a
        noise-averaged NLIF as a reward function. Both choices are investigated
        in the subsequent section.

    \section{Results}\label{sec:results}
    In all cases below, we optimise for a CNOT gate, as we would like to produce
    accurate entangling protocols as a general requirement for useful quantum
    algorithms.

    An RL loop consists of inputting an action (\(a_{t}\)), and obtaining
    feedback (\(o_{t}, r_{t}\)). In all of the scenarios below, we keep the
    former as a number in a continuous bounded interval. We investigate the
    nuances of RL agent design components with respect to the observation and
    reward functions, and how they affect agent performance. First, the effects
    of finite rise-time are investigated (\cref{ssec:resobs}), followed by the
    effects of noise in different reward setups (\cref{ssec:nevo},
    \cref{ssec:robust_exact}). Finally, we examine a setup with tomography
    statistics (\cref{ssec:robust_pt}).

    Additionally, in the following experiments, we train over multiple seeds to
    take into account the stochasticity of neural network parameter
    initialisation, the randomness of RL's exploration-interaction based
    training, and the randomness in the cases of quantum dot system noise as
    well as the simulated shot noise of gate tomography. For noise-averaged
    fidelities (\cref{ssec:robust_exact}, \cref{ssec:robust_pt}), we take the
    maximum over our training realisations as we are only interested in
    obtaining the best possible pulses.
    \subsection{Observations}\label{ssec:resobs}
        Even in the simplest regime of no shot or system noise, the delayed
        response requires attention, as it presents a challenge to the
        Markovianity of the decision process. For unitary-only based
        observations, if the sampling interval \(T_{s}\) is shorter than the
        delay caused by the finite rise-time effects, the effect of the previous
        action is not fully accounted for, as the agent must choose another
        action before the full effects of previous action are realised.

        A simple solution is to fix this rate so this situation is avoided.
        However, too few actions in a chosen protocol length can affect the
        controllability, impacting performance. We investigate the inclusion of
        the current value of the pulse amplitudes in the observation model as a
        method to inform the agent of the achievable bounds of its current
        convoluted action. This restores the Markovianity of the setup. In the
        case of a greater delay (i.e.\@ greater than \(2T_{s}\)), the set of
        \(n\geq2\) previous pulses would be required. We test this hypothesis by
        comparing the agent's performance when the observation is \(o_{t} = (T -
        t, U_{t})\) and \(o_{t} = (T - t, \epsilon_{t}, U_{t})\), where
        \(U_{t}\) corresponds to an exact tomography of the gate.

        We use a simple reward function to measure agent performance. We use a
        sparse reward based on the final negative log-infidelity between the
        produced gate and the CNOT gate:
        \begin{equation}\label{eq:sprsrwrd}
            r_{t} = \begin{cases}
                f & \text{if } t = T\\
                0 & \text{otherwise}
            \end{cases}\\
        \end{equation}

        For simplicity, the actual response kernel (\cref{fig:qd_impulse} (b))
        is replaced with a plain Gaussian function with a fitted standard
        deviation to the actual response function. We take 50 actions with a
        protocol length of 50ns, such that the sampling period is \(T_{s} =
        1\text{ns}\). We manipulate the peak position of the Gaussian to produce
        smaller or larger delay effects on the input pulse.

\begin{figure}
            \centering
            \includegraphics[scale=1.0]{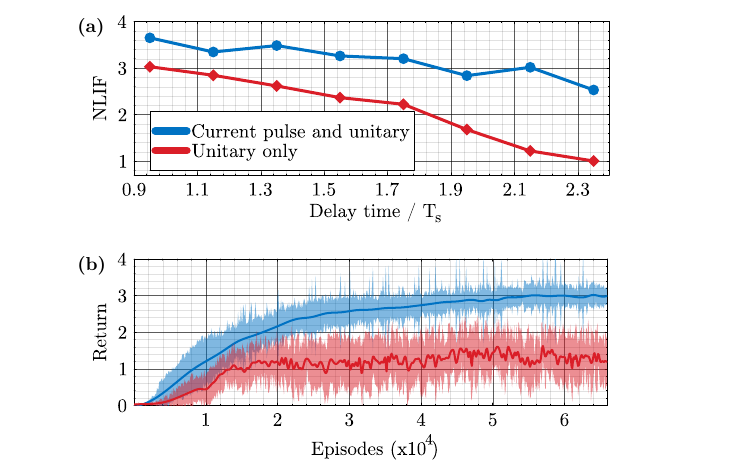}
            \caption{%
                (a) A plot showing the effect of the delay time (divided by the
                sampling period) of a convolution filter (i.e. the mean of a
                Gaussian kernel) with respect to the final mean NLIF (between
                our RL-produced gate and CNOT) obtained by an agent after
                training on a noise-free quantum environment (over many training
                seeds). In the case \(o_{t} = (T - t, U_{t})\) (\(U_{t}\) is an
                exact tomography at each time step), we see a degradation of
                performance as the mean of a Gaussian kernel is shifted in time,
                corresponding to decrease in the Markovianity of the situation.
                Inclusion of the current pulse in agent's observation (i.e.
                \(o_{t} = (T - t, \epsilon_{t}, U_{t})\)) leads to a drastic
                improvement in performance. (b) The training curves (return
                (cumulative reward) vs. episodes) of the two choices using a
                kernel with a delay time of \(\sim2.15\)ns. The agent learns
                superior protocols faster and with less variance over different
                seeds (shaded) when the extra information about the pulse is
                included in the observation.
            }\label{fig:observation_comparison}
        \end{figure}

        The comparative performance between the two observation models is shown
        in~\cref{fig:observation_comparison} (a). We see a drastic improvement
        of up to two orders of magnitude in the agent's performance as delay
        time relative to the sampling period increases. In addition to the
        greater reward, the stochasticity of the training is reduced, signifying
        a greater ease of learning, as seen in~\cref{fig:observation_comparison}
        (b). Including more of the pulse history may be required if the delay is
        greater than two actions, as this would restore the Markovian properties
        and allow the agent to make better informed decisions. This choice is
        maintained in the following sections.
    \subsection{Noisy evolution}\label{ssec:nevo}
        We will now examine the agent's performance in the full presence of all
        noise contributions (\cref{eq:noise}) and an observation that includes
        the current pulse, necessitated from the above discussion. In the
        presence of system noise, we simply use the noisy unitary evolution
        \(\tilde{U_{t}}\) (i.e. the unitary resulting from a particular noise
        realisation) as part of our agent's observation input, such that the
        total observation is \(o_{t} = (T - t, \epsilon_{t}, \tilde{U_{t}})\).
        As before, a sparse reward (\cref{eq:sprsrwrd}) is employed, and we
        investigate agent behaviour with noisy trajectories generated by
        hyperfine noise on the drift and fast and slow charge noise
        contributions on the pulses.

        We evaluate the performance of an agent across a range of total protocol
        lengths and number of actions (i.e. inputs to the AWG), and examine
        performance variability, aiming to achieve a desired performance of a
        fidelity of at least \(\mathcal{F}=0.999\).

\begin{figure}
            \centering
            \includegraphics[scale=0.25]{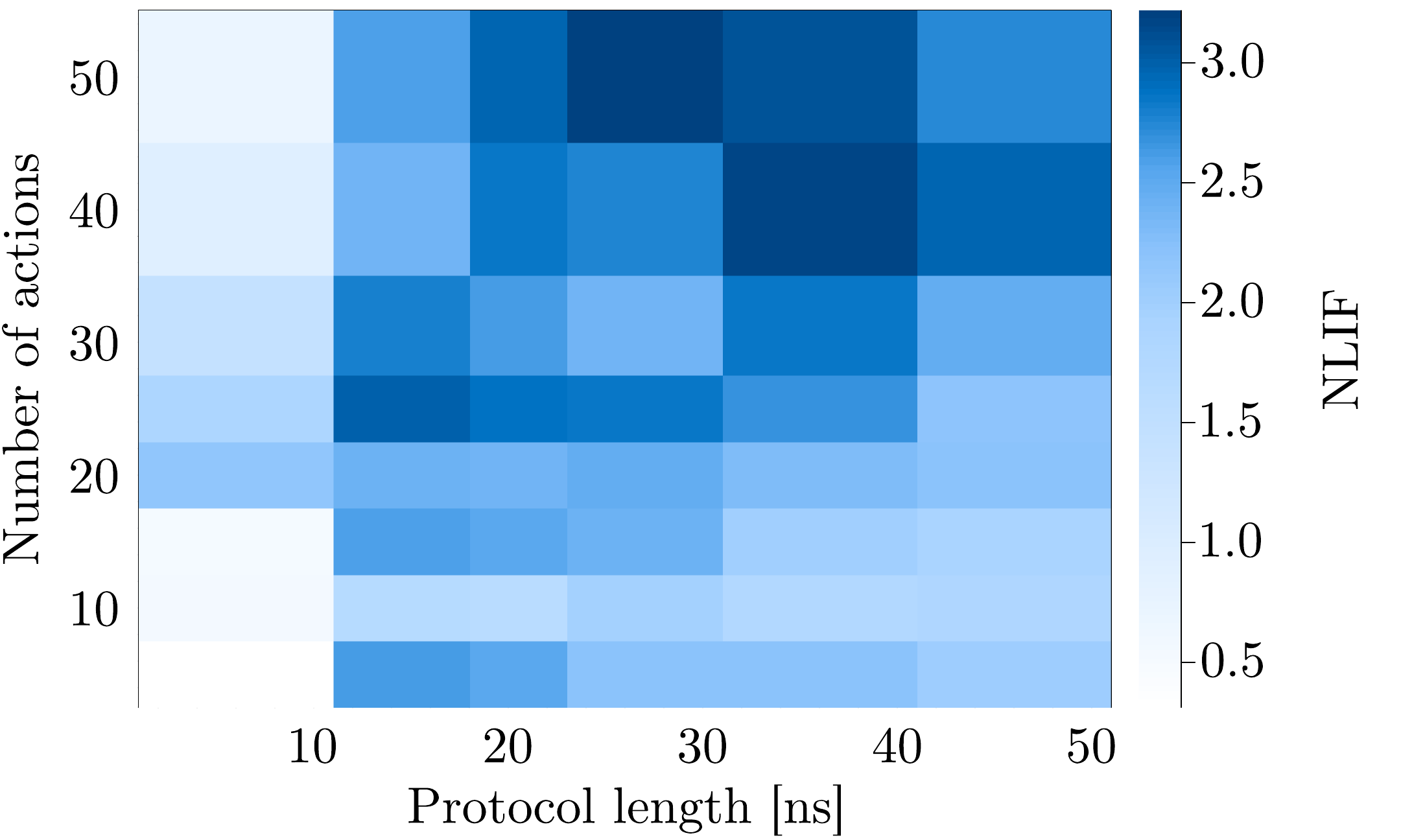}
            \caption{%
                A heatmap of average final negative log fidelities with varying
                number of actions and protocol lengths. There are some sweet
                spots with superior performance, but the agent achieves NLIFs of
                \(> 3\) in multiple scenarios, with universally poor performance
                past the quantum speed limit.
            }\label{fig:rewards_noisy_evolution}
        \end{figure}

        From~\cref{fig:rewards_noisy_evolution}, we observe some key features.
        Firstly, the quantum speed limit is manifest when the total protocol
        length is short, resulting in poor performance across all possible
        number of actions. Secondly, although there are multiple regions that
        exceed the performance threshold, there is variability in the
        combination of protocol length and number of actions. We identify
        sweetspots where fidelities are greater, with a preference for
        combinations involving shorter protocol lengths and a fewer actions.

\begin{figure}
            \centering
            \includegraphics[scale=1.0]{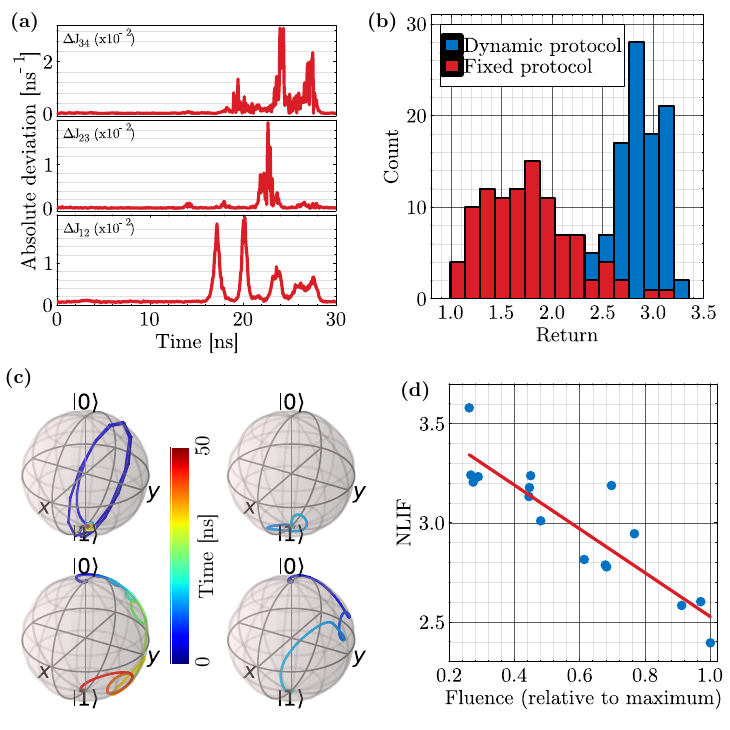}
            \caption{%
                (a) The absolute deviation between two different exchange
                interactions of post-filtered protocols, each with different
                noise realisations. The agent is able to learn adaptive control
                protocols, with different noise realisations leading to slight
                adjustments in the protocol to maintain performance. (b) Using
                the same protocol for subsequent noise realisations results in
                poorer performance (red), whereas using a dynamical protocol
                greater performance (blue). (c) Time evolution on the separable
                parts of the Bloch sphere for each logical qubit starting from
                \(\ket{10}\) for different protocol lengths and actions with (20
                actions, 20 ns) (right) and (40 actions, 50 ns) (left). We see
                the trajectories vary demonstrating the ability of RL agents to
                find a diverse set of solutions. (d) The relative fluence
                (cumulative power of pulses) vs. the final mean reward. Pulses
                that produce higher fidelities also minimise the total spent
                energy.
            }\label{fig:noisy_evolution}
        \end{figure}

        One artefact of RL optimisation is its adaptability to fluctuations of
        the environment, a feature of the conditional distribution-based
        training of RL. We investigate this aspect by considering how a trained
        agent's protocols change in different noise realisations. As we see
        in~\cref{fig:noisy_evolution} (a), two different noise realisations lead
        to minor adjustments in the agent's actions, resulting in variations in
        the final protocol (exchange interactions) that maintain optimal
        performance of \(\mathcal{F}=0.999\). or greater. On the contrary, if
        the protocol was fixed, the performance over differing noise
        realisations may vary significantly\@.
        Additionally,~\cref{fig:noisy_evolution} (b) shows the variability in
        performance when using the best-performing fixed protocol compared to a
        dynamic, noise-dependent adjusting protocol, consisting of a sequence of
        20 steps and with a protocol length of 30 ns, over 100 noise
        realisations. The latter maintains a mean infidelity of approximately
        \(10^{-3}\). An alternative strategy would be to optimise with respect
        to a noise-averaged reward in order to produce protocols that are
        performative over the different realisations. This method is employed in
        the next section.

        We also explore how the different choices of protocol length and number
        of actions affect the trajectory of the separable part of the Hilbert
        space. We apply our CNOT gate to \(\ket{10}\) (i.e. a controlled bit
        flip) with two different combinations of protocol length and number of
        actions. In~\cref{fig:noisy_evolution} (c), for a shorter protocol and
        fewer actions, we observe that the target qubit transitions faster
        towards the opposite pole of the Bloch sphere. However, it is still
        over-parameterised, resulting in non-geodesic behaviour. On the
        contrary, a longer protocol with many actions has periods of low pulses,
        leading to circular revolutions on the Bloch sphere. This is optimal
        behaviour as minimal energy is applied and the fluence (i.e. the
        integral over time of pulse power) is at a minimum. Indeed, as
        demonstrated in~\cref{fig:noisy_evolution} (d), protocols that have
        higher mean reward at the end of training also have a lower total power.
        This illustrates that RL's open-box optimisation may lead to drastically
        different physical behaviour, including desirable secondary outcomes.
    \subsection{Noise-robust entangling strategies}\label{ssec:robust_exact}
        For a more realistic scenario closer to a feasible experimental setup,
        we aim to find fixed control protocols, that are robust against
        fluctuations in system parameters induced by noise effects, but without
        the constant monitoring of noisy trajectories as in the previous
        section.

        In this setting, the choice of observation is the time-to-go, current
        pulse, and a function of the simulated unitary gate in the absence of
        any noise. Although the true evolution of the system is noisy, this
        observation constitutes a belief state and corresponds to the
        expectation of noisy trajectory realisations. In other words, we can
        perform experimentally realisable closed-loop control by providing the
        agent with information about the expected trajectory of its actions,
        without direct input about the current noise realisation.

        Additionally, since there is a (non-injective) mapping from the pulse
        history to the current unitary operator, an observation of either the
        gate or the pulse history can be used (i.e. \(o_{t} = (T - t,
        \epsilon_{t}, \epsilon_{0:t})\) or \(o_{t} = (T - t, \epsilon_{t},
        U_{t})\)). We investigate the performance of the agent with respect to
        these two choices of observation.

        To find robust protocols with respect to system noise, an averaged
        infidelity is used, whereby an agent optimally finds a protocol
        that works across many different noise realisations:
        \begin{equation}\label{eq:rbstrwrd}
            r_{t} = \begin{cases}
                {\left\langle f\right\rangle}_{N_{r}}
                = \frac{1}{N_{r}}\sum^{N_{r}}_{n} f_{n} & \text{if } t = T\\
                0 & \text{otherwise}
            \end{cases}
        \end{equation}
        Here, \(N_{r}\) is the number of realisations, chosen such that there is
        a convergence in agent performance. For each realisation, the unitary
        gate used in the fidelity \(f_{n}\) is the final noisy gate obtained
        using the same pulses generated by the agent, but with different system
        noise contributions.

        In this regime, our experimental conditions are the same as for
        traditional noise-robust, gradient-based methods, absent of tomographic
        data for the unitary, which would only be performed in the final time
        step to calculate the fidelity. Consequently, the resulting performance
        can be suitably compared with previous results. Cerfontaine~\textit{et
        al.}~\cite{Cerfontaine_2020} reported a fidelity past the threshold of
        \(99.9\%\) using their gradient-based approach mentioned mentioned
        in~\cref{sec:qd}. We use their method as implemented in the qopt
        library~\cite{Teske_2022} as a competitive benchmark.

        We note that the fidelities reported below lie lower than said
        reference, indicating that the two models are not exactly equivalent.
        Further analysis revealed, that slightly adjusted sample periods render
        comparable reachable fidelities, as we will discuss towards the end of
        this section. We optimised over a set of protocol lengths and number of
        actions in both approaches.

\begin{figure}
            \centering
            \includegraphics[scale=1.0]{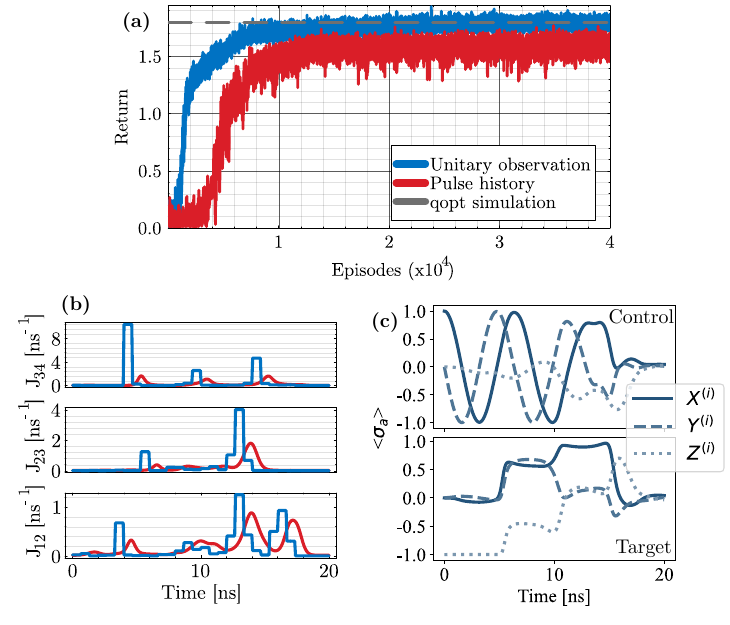}
            \caption{%
                (a) The training curves for an RL agent with a robust reward
                consisting of maximum negative log infidelities. The RL agent
                will perform marginally better if it has full access to the
                current unitary gate at time step \(t\) compared to only having
                the history of actions (with which a non-injective mapping to a
                unitary can be constructed). The former has a similar
                performance to our gradient based-simulation using qopt (only
                consisting of the drift noise term). (b) A full protocol of a
                robust pulse of 20 ns with 30 actions. In blue are the actions
                corresponding to the discretised pulses inputted to the AWG
                device and the red lines are the smoothed pulses affected by the
                finite response of the qubit. (c) The expectation values of the
                logical qubits for an entangling operation on \(\ket{0+}\) with
                respect to time. The logical qubit (below) is entangled in steps
                with a sequence of rotations on the physical qubits.
            }\label{fig:robust_reward_pulses}
        \end{figure}

        \cref{fig:robust_reward_pulses} (a) shows that the agent can learn to an
        NLIF of about 1.8 in this regime with full observation of the noise-free
        unitary at all but terminal time steps and to an NLIF of about 1.5 with
        just the pulse history, which requires significantly less information.
        In the former case, the performance is similar to our qopt simulation
        (that only consists of drift noise). The protocol duration in this
        setting is 20 ns and consists of 30 actions.

        However, we note that this performance does not appear to stem from
        limitations of RL training with system noise. First, we simplify the RL
        environment to a simple one-qubit quantum system, with one control (and
        exchange interaction) and drift term, and only use the drift noise as it
        is the dominating terms for the infidelity at this sampling period.
        In~\hyperref[ssec:1dq]{Appendix D}, we show that, as we scale the noise,
        both RL and qopt fidelities scale equivalently. Therefore there may be a
        fundamental upper bound on the averaged fidelity with noisy time
        evolution. Likewise, training curves yield similar infidelity limits
        that match qopt.

        A typical obtained protocol (\cref{fig:robust_reward_pulses} (b)) has
        hard kicks in the input signals or actions, causing weaker smoothened
        signals that affect the exchange interactions between the four dots.
        First, the two logical qubits are rotated locally by \(J_{12}\) and
        \(J_{34}\), such that they can be suitably entangled by \(J_{23}\), and
        then they are rotated back to the original basis. This can also be seen
        in the expecation values of each Pauli operator on each logical qubit
        in~\cref{fig:robust_reward_pulses} (c), when starting with an initial
        state of \(\ket{+0}\), which allows for entanglement.

\begin{figure}
            \centering
            \includegraphics[scale=1.0]{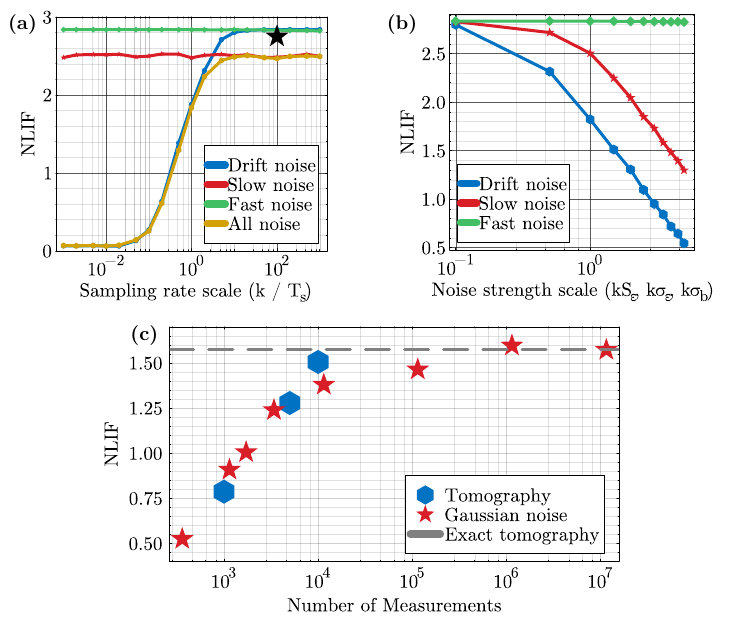}
            \caption{%
                (a) The infidelity scaling of the noise contributions with
                respect to the frequency rate (i.e. \(k / T_{s}\) for scale
                factor \(k\)) without reoptimisation. We observe that the
                hyperfine noise contributions dominate in low frequency scales,
                whilst the slow charge noise is independent. The overall factor
                of the fast charge noise makes this noise contribution
                negligible. The black star is the fidelity achieved from a newly
                trained agent trained from scratch at a scale of 100. We observe
                that its performance surpasses the expected baseline. (b)
                Infidelity scaling with respect to the noise scale (i.e.
                \(k(S_{\epsilon} | \sigma_{\epsilon} | \sigma_{b}\))).  While
                the fast charge noise is dominated mostly by the time scale and
                therefore suppressed, the hyperfine and slow charge noise scales
                exponentially. In other words, small increases in the noise
                strengths have negligible effect on agent performance. (c)
                Negative log-infidelity as a function of the number of
                measurements per episode used to perform process tomography. In
                circles are unitaries reconstructed by POVM-based multinomial
                statistics, while the stars represent an approximation using
                additive Gaussian noise with a standard deviation scaled to
                match the expected number of measurements to reach the same
                infidelity with POVM-tomography (\hyperref[ssec:u_t]{Appendix
                C}). The final NLIF converges to the exact tomographic
                performance as the number of measurements becomes large.
            }\label{fig:robustness_scaling}
        \end{figure}

        To investigate the effect of different noise contributions based on the
        sampling periods \(T_{s}\), we examine the scaling of the individual
        noise contributions by adjusting the time and energy scales
        (reciprocally, thus keeping the overall evolution duration fixed) of the
        system but reuse the same obtained protocol without
        re-optimisation~\cite{Cerfontaine_2020}. We then plot the infidelity
        change of individual noise contributions (i.e. the others are switched
        off) as well as the overall noise with respect to the scaling of the sample
        period (i.e. \(k / T_{s}\) where \(k\) is the scale factor)
        (\cref{fig:robustness_scaling} (a)). We observe that regions of long
        sample periods (and therefore less frequent qubit manipulation) are
        heavily affected by qubit drift or hyperfine noise. At lower sampling
        periods, we see a negligible rise in the noise contributions of the fast
        time-correlated charge noise, and it is expected to further scale with
        respect to the power \(\alpha\), but is otherwise suppressed by the
        overall scale. We conclude that choosing a higher sampling rate by an
        order of magnitude may be prudent to get the most out of agent
        performance. To confirm this, we train an agent from scratch at a scale
        factor of \(100 / T_{s}\) and observe that its performance surpasses the
        expected baseline from the sampling period scaling.

        In addition, we also evaluate the robustness of our protocol as we
        increase the relative noise strengths in~\cref{fig:robustness_scaling}
        (b). While the fast charge noise is dominated mostly by the time scale
        and therefore suppressed, the hyperfine and slow charge noise scales
        exponentially past the initial baseline value, which we can see by the
        linear relation in the log-log plot.
    \subsection{Process Tomography}\label{ssec:robust_pt}
        Finally, for the most complex case, we slightly modify the previous
        setup to make it as close as possible to a feasible experimental setup.
        For the calculation of fidelity, our obtained unitary is estimated using
        limited or partial tomography and its associated measurement statistics
        and shot noise. We investigate the effects between taking a small and a
        large number of measurements on agent performance due to this added
        uncertainty. In other words, we explore whether shot noise causes
        confusion to the agent.

        To this end, we take the following approach: First, using a fixed pulse
        generated by the agent, we obtain a unitary after an evolution with a
        system noise realisation. Then, we perform a single-shot measurement of
        this unitary by selecting a random input quantum state, as process
        tomography requires a suitable set of state tomography experiments, as
        well as a random POVM element (\hyperref[ssec:u_t]{Appendix C}). We then
        measure the outcome of this combination of POVM element and input state
        and only one data point is collected per system noise realisation. We
        repeat this process for \(N_{m}\) runs (i.e. we collect \(N_{m}\)
        snapshots). Using the collected measurement statistics, we then
        reconstruct a unitary which includes the statistics of both system and
        shot noise. This unitary is then used for calculating the fidelity, and
        hence the reward.

        However, due to the high computational cost of our experiments, due to
        performing a large number of snapshots multiplied by the requirement of
        training over many episodes, we also apply an alternative approach. We
        revert to the setup in the previous section, but for each unitary
        generated by a realisation of system noise, we add an additional
        additive Gaussian noise. In other words, we model the shot noise using a
        Gaussian with a chosen standard deviation:
        \begin{equation}
            r_{T} = {\left\langle
                f\left(\text{CNOT}, U_{T} + \delta U\right)
            \right\rangle}_{%
                \delta U
                \sim
                \mathcal{N}\left(
                    \mathbf{0}, \mathbf{\sigma^{2}\mathds{1}}
                \right)
            }
        \end{equation}
        Here, \(U_{T}\) is a noisy realisation of the generated unitary and we
        add to it a \(d\times d\) noisy matrix, representing the shot noise.

        To ensure the accuracy between the previous POVM-based approach and our
        alternative simulated measurement noise, we match the infidelity scaling
        (of only the shot noise and ignoring the system noise) of the additive
        Gaussian noise with standard deviation \(\sigma\) with the infidelity
        scaling of a POVM-based approach using a certain number of snapshots. In
        other words, we can crudely approximate the number of snapshots \(N\) to
        a standard deviation or noise scale (\hyperref[ssec:u_t]{Appendix C}).

        In~\cref{fig:robustness_scaling} (c), we observe the effect of the
        number of snapshots per episode on the final NLIF in a regime consisting
        of 30 actions and a protocol length of 30 ns. There is a near
        order-of-magnitude gain of performance as the number of snapshots per
        episode increases, converging to the exact tomographic result based on
        the previous subsection. Notably, by taking just 1000 snapshots, the
        agent converges to an NLIF of approximately 0.8. However, this result is
        not due to limitations in  agent learning, but rather the fidelity bound
        of an estimated unitary gate with 1000 snapshots, as can be seen
        in~\hyperref[ssec:u_t]{Appendix C}. Furthermore, the tomographic data
        matches the alternative Gaussian approach for a smaller number of
        measurements, such that we can reliably infer the expected tomographic
        performance in the large measurement limit without the computational
        constraints. Consequently, our experimentally realistic protocol is
        constrained by the desired fidelity threshold, which may be bounded by
        the shot noise if insufficient measurements are performed, but is
        ultimately limited by the system noise. In other words, shot noise does
        not lead to additional confusion in agent learning.

    \section{Conclusion}
    We investigated the feasibility of using RL to control realistically
    modelled quantum hardware, subject to three major constraints: rise-time
    effects, multiple noise contributions to the system (including
    time-correlated noise), and limited information about the realised unitary
    via simulated tomography in the presence of the aforementioned system noise.
    We sequentially added each constraint to test RL-agent performance in the
    face of increasing complexity.

    Firstly, we demonstrated that proper and carefully considered agent design
    is necessary to achieve satisfactory performance, even if seemingly trivial.
    Small changes in RL state observation had drastic effects on the final gate
    fidelities achieved, as the finite rise-time effects of controlling hardware
    broke the Markovian properties required needed for successful learning.
    Including pulse information in the observation restored these properties.

    Secondly, we introduced decohering noise to the quantum system, but allowed
    the agent to observe deviations in the unitary trajectory. We achieved a
    performance of 99.9\% fidelity for certain combinations of protocol lengths
    and number of agent actions. Additionally, in this setting, the agent varies
    its inputs slightly to correct for the noise to maintain an optimal
    performance. Furthermore, optimal pulses were found to minimise the total
    power as a nascent outcome.

    Thirdly, we matched or surpassed the baseline performance of our
    alternative, gradient-based approach in similar regimes by using a
    noise-averaged reward to find a single noise-robust pulse. Furthermore,
    given only information about the protocol (i.e. sequence of pulses), an
    agent is able to infer a mapping onto a unitary and perform close to the
    baseline without direct information of the unitary (sans a final fidelity
    reward function), and is subsequently free of any model-bias.

    Finally, we performed a simulated process tomography on the generated
    unitary to calculate the fidelity. This meant that the agent had to learn in
    the presence of both system and measurement noise. Despite the additional
    shot noise, we found that the agent performance was constrained by the
    number of measurements that matched the fidelity of quantum process
    tomography alone. In other words, with enough measurements, achieved
    fidelities were only limited by the system noise and shot noise did not
    contribute to a greater confusion in learning.

    Two avenues would be of interest for future research, independent of the
    specific caveats of RL control of quantum dots. On the side of RL design, it
    would be interesting to investigate the effect of incorporating
    autoregressive neural network structures to augment agent capabilities,
    particularly when only minimal information is available, such as with
    pulse-only observation data provided to the agent.

    Another avenue of interest would be to investigate incorporating RL into
    existing frameworks. One advantage of using RL with a noise-averaged
    fidelity as a reward function is that this training can be used as a form of
    offline pre-training with a known Hamiltonian (and noise) model.
    Consequently, once a simulated, noise-averaged baseline has been
    established, it is possible to further optimise the protocol with respect to
    real-device noise. This additionally avoids the inference overhead of neural
    networks, with an alternating closed-loop sequence of many runs and their
    respective fidelities and a RL-training step to maintain or improve
    noise-robust pulses. This pre-training stage could easily be replaced with
    existing gradient-based or model-aware methods, with RL being used to
    improve them yet remaining general enough to train from scratch.
\section*{Acknowledgements and data availability}
    We thank Max Oberländer for helping us with using the qopt package developed
    at PGI-11 and Hendrik Bluhm for helpful comments on the manuscript. This
    work was supported through the Helmholtz Initiative and Networking Fund,
    Grant No.~VH-NG-1711.

    The code used for this paper was made by the authors in the Julia
    programming language~\cite{Julia-2017} and is in active development at
    \hyperlink{https://github.com/EmergentSpaceTime/RLQuantumControl}{%
        https://github.com/EmergentSpaceTime/RLQuantumControl}. The RL generated
    data which was used for this paper as well as the scripts to generate the
    plots can be found
    at~\hyperlink{10.5281/zenodo.16900638}{10.5281/zenodo.16900638}.

    \appendix
    \section{Reinforcement Learning}\label{ssec:rl_appendix}
        Readers interested in an in-depth description of reinforcement learning
        are referred to Ref.~\cite{sutton2018reinforcement}. The following is a
        brief overview of the main concepts. Reinforcement learning (RL) is
        built around three primary components: observation, decision-making, and
        feedback. The central framework uniting these aspects are \textit{MDPs}.
        \begin{definition*}\label{def:mdp}
            A \textbf{Markov decision process} (MDP) is the tuple {%
                \(
                    (
                        \mathcal{S},\
                        \mathcal{A},\
                        \mathcal{T},\
                        \mathcal{R},\
                        \gamma
                    )
                \)
            } where:
            \begin{itemize}
                \item \(\mathcal{S}\) is a set of states
                \item \(\mathcal{A}\) is a set of actions
                \item {%
                    \(
                        \mathcal{T}:
                        \mathcal{S} \times \mathcal{S} \times \mathcal{A}
                        \longrightarrow [0, 1]
                    \)
                    is the state-transition probability:\\
                    \(
                        T^{a}_{ss'} = {%
                            \mathds{P}\left[
                                s_{t + 1} = s' | s_{t} = s, a_{t} = a
                            \right]
                        }
                    \)
                }
                \item {%
                    \(
                        \mathcal{R}:
                        \mathcal{S}\times\mathcal{A}
                        \longrightarrow \mathds{R}
                    \)
                    is the reward function:\\
                    \(
                        r^{a}_{s} = {
                            \mathds{E}[r_{t + 1}|s_{t} = s, a_{t} = a]
                        }
                    \)
                }
                \item \(\gamma \in [0, 1]\) is the discount factor
            \end{itemize}
            An MDP has the Markovian property of independence from past and
            future states.
        \end{definition*}
        Additionally, to formalise decision-making, we must define a
        \textit{policy}.
        \begin{definition*}
            A \textbf{policy} (\(\pi \)) is a conditional probability
            distribution over the set of actions \(\mathcal{A}\) and the set of
            states \(\mathcal{S}\), denoted \(\pi(a|s)\).
        \end{definition*}
        A set of states, \(\mathcal{S}\), represents the possible configurations
        of an \textit{environment}. An environment can be anything from a chess
        game to a model of urban traffic. An \textit{agent} is an abstract
        object that performs an \textit{action} from a set of possible actions,
        \(\mathcal{A}\), which affect the state of the environment. Once an
        action is performed, the state of the environment transitions to another
        possible state via a conditional probability function. For example,
        going from any one chess configuration to another may not be possible.
        In other words, the transition function models the dynamics of the
        environment.

        Decision-making cannot be informed by states and actions alone. To
        achieve this, an objective or goal is required. Such an objective is
        modelled with a \textit{reward}, enabling the possibility of
        optimisation or improvement. To model an agent's choice of action or its
        decision-making a policy is used. The goal of RL is to find a optimal
        policy \(\pi^{\star} \) that maximizes the cumulative reward:
        \begin{definition*}
            The \textbf{(discounted) return} \(G_{t}\) is the cumulative
            reward obtained starting from the time step \(t\):
            \begin{equation}
                G_{t} = \sum_{k=0}^{\infty}\gamma^{k}r_{t + k + 1}
            \end{equation}
        \end{definition*}
        A policy is evaluated through the following function:
        \begin{definition*}
            The \textbf{action-value function} or \textbf{q-function} is the
            expected return starting from a state \(s\), taking taking an action
            \(a\), then following a policy \(\pi \):
            \begin{equation}
                q_{\pi}(s, a) = \mathds{E}_{\pi}{%
                    [G_{t}|s_{t} = s, a_{t} = a]
                }
            \end{equation}
        \end{definition*}
        These q-functions have the following optimality condition:
        \begin{equation}
                q^{\star}_{\pi} = {%
                    \underset{\pi}{\text{max}}\ q_{\pi}(s, a)
                }
        \end{equation}
        The optimal policy is not unique, but its q-function is. We can use the
        recursive nature of the above equations to find an optimal
        solution via dynamic programming, giving the Bellman equation:
        \begin{equation}\label{eq:bellman}
            q^{\star}_{\pi}(s, a) = {%
                r^{a}_{s}
                + \gamma
                \sum_{s'\in\mathcal{S}}T^{a}_{ss'}
                \underset{a'}{\text{max}}\ q^{\star}(s', a')
            }
        \end{equation}

        Two important challenges arise from using~\textit{tabular} methods (i.e.
        keeping a memory of all combinations of state and actions to calculate
        the q-function). Firstly, many MDPs have state and actions spaces that
        are intractable. For example, the game of Go has \(~10^{170}\)
        configurations. Furthermore, MDPs with continuous states and actions
        cannot be modelled. Secondly, in~\cref{eq:bellman}, we explicitly need
        the dynamics of the system via the transition probabilities, which are
        often either unavailable or intractable, as previously aformentioned.

        For the first problem, since the state and action spaces of many control
        problems, including quantum control, are continuous or involve vast
        state spaces, we parameterise our Q-functions and policies using neural
        networks. They satisfy the universal function approximation
        theorem~\cite{ferrari2005smooth}, therefore can approximate any function
        up to local convergence. This allows us to construct a function mapping
        continuous or discrete state and actions to a q-value or policy
        probability distribution.

        For the second problem, we employ temporal-difference (TD) learning.
        This approach involves maintaining an improving estimate of the
        q-function. Learning is combined with gaining experience, as this
        action-value-estimate is updated in an online fashion. Initially, after
        each time step, we calculate a target towards which we move our current
        q-estimate. This target itself is an estimate of the current q-function.
        In other words, we update a guess towards a guess. From the recursive
        nature of the Bellman equation, we aim to converge to the true
        q-function by minimising the following error:
        \begin{equation}\label{eq:td_error}
            \delta_{t} := {%
                r_{t + 1}
                + \gamma\
                \underset{a}{\text{max}}\ q(s_{t + 1}, a)
                - q(s_{t}, a_{t})
            }
        \end{equation}

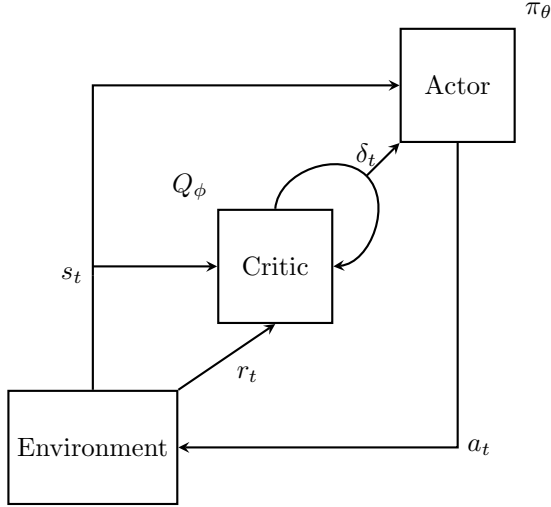
\begin{figure}
            \centering
            \begin{tikzpicture}[scale=0.6]
                {%
                    \node[
                        rectangle,
                        draw,
                        thick,
                        minimum size=15mm,
                        label=above right:\(\pi_{\theta}\),
                    ] (A)
                    at (4, 4)
                    {Actor};
                }
                {%
                    \node[
                        rectangle,
                        draw,
                        thick,
                        minimum size=15mm,
                        label=above left:\(Q_{\phi}\),
                    ] (C)
                    at (0, 0)
                    {Critic};
                }
                {%
                    \node[rectangle, draw, thick, minimum size=15mm] (E)
                    at (-4, -4)
                    {Environment};
                }
                \node (N) at (-4, 0) {};
                \node (AC) at (2, 2) {};
                {%
                    \draw[thick, ->, >=stealth]
                    (A.south) |-
                    node [midway, right] {\(a_{t}\)}
                    (E.east);
                }
                {%
                    \draw[thick]
                    (E.north) to
                    node [at end, left] {\(s_{t}\)}
                    (N);
                }
                \draw[thick, ->, >=stealth] (N.south) |- (A.west);
                \draw[thick, ->, >=stealth] (N.center) -- (C.west);
                {%
                    \draw[thick, ->, >=stealth]
                    (E.north east) -- node[midway, below right]
                    {\(r_{t}\)}
                    (C.south);
                }
                \draw[thick, bend left=65] (C.north) to (AC.center);
                {%
                    \draw[thick, ->, >=stealth, bend left=65]
                    (AC.center)
                    to node[at start, above] {\(\delta_{t}\)}
                    (C.east);
                }
                \draw[thick, ->, >=stealth] (AC.center) to (A.south west);
            \end{tikzpicture}
            \caption{%
                A general overview of actor-critic methods in RL\@. The critic
                (i.e. a parameterised q-function) judges the actor (i.e. a
                parameterised policy) and gives feedback whilst simultaneously
                improving its estimation using the temporal difference error
                \(\delta_{t}\) (\cref{eq:td_error}).
            }\label{fig:actor_critic}
        \end{figure}

        Combining the two solutions, we employ \textit{actor-critic} style
        methods (\cref{fig:actor_critic}), simultaneously optimising
        parameterised policies and parameterised q-functions using gradient
        descent methods such as ADAM~\cite{kingma2014adam}. An \textit{actor} is
        nothing more than a parameterised policy, which takes a state and
        outputs a conditional distribution of possible actions. A
        \textit{critic} is a parameterised q-function which takes a state-action
        pair and outputs a number.

        In addition, we employ a chosen algorithmic variation of actor-critic
        known as soft actor-critic~\cite{haarnoja2018soft, haarnoja2019soft}.
        Here, we augment the reward with the current policy entropy to encourage
        exploration:
        \begin{equation}
            r'_{t} = {%
                r(s_{t}, a_{t}) + \alpha H(\pi_{\theta}(\cdot|s_{t}))
            }
        \end{equation}
        A controlled temperature parameter, \(\alpha \), is used to interpolate
        between high and low exploratory behaviour. The recursive Bellman update
        becomes:
        \begin{equation}
            \begin{split}
                Q_{\phi}(s_{t}, a_{t}) &= r_{t}\\
                &+ \gamma Q_{\phi}(s_{t + 1}, a_{\text{new}}(s_{t + 1}))\\
                &- \alpha\log{\pi_{\theta}(a_{\text{new}}(s_{t + 1})|s_{t + 1})}
            \end{split}
        \end{equation}
        Here, \(a_{\text{new}}(s_{t + 1}) \sim \pi_{\theta}(\cdot |s_{t + 1})\).
        The q-network is then trained to minimise this modified temporal
        difference leading to (after some algebraic manipulation), a cost
        function for the q-network:
        \begin{equation}
            J_{\phi} = {%
                \frac{1}{|\mathcal{B}|}
                \sum_{\tau\sim\mathcal{B}}{\left(
                    Q_{\phi}(s, a) - \delta(s, r')
                \right)}^{2}
            }\\
        \end{equation}
        Here, \(\delta(r, s')\) is defined to be:
        \begin{equation}
            \delta(r, s') = {%
                r
                + \gamma\left(
                    Q_{\phi}(s', a_{\text{new}}(s'))
                    - \alpha\log{\pi_{\theta}(a_{\text{new}}(s')|s')}
                \right)
            }
        \end{equation}
        Here, \(\mathcal{B}\) refers to batches of sampled experience
        trajectories, with \(\tau = (s, a, s', r)\). The policy cost function is
        to maximise the expected return:
        \begin{equation}\label{eq:policy_cost}
            J_{\pi}(\theta) = \underset{\tau\sim\pi}{\mathds{E}}\left[
                G(\tau)
            \right]
        \end{equation}
        Since the rewards are estimated using the q-function, we can rewrite
        this as:
        \begin{equation}
            J_{\theta} = {%
            \frac{1}{|\mathcal{B}|}\sum_{\tau\sim\mathcal{B}}
            \left[
                -\alpha\log{\pi_{\theta}(\tilde{a}_{\theta}(s)|s)}
                + Q(\tilde{a}_{\theta}(s), s)
            \right]
        }
        \end{equation}
        Note that the policy loss requires sampling new actions \(a_{\theta}(s)
        \sim \pi(\cdot|s)\), which are obtained using the reparametrisation
        trick~\cite{kingma2022autoencoding} in order to compute gradients.

        Further enhancements to SAC include using a distribution of critics and
        taking the minimum of the q-values to reduce overestimation
        bias~\cite{kuznetsov2020controlling}. Furthermore, the q-networks may be
        augmented with dropout layers that have empirically demonstrated
        improving sample
        efficiency~\cite{hiraoka2022dropoutqfunctionsdoublyefficient}.

        Finally, an important aspect of MDPs must be mentioned. In this
        appendix\footnote{And other parts of this paper.}, we have
        interchangeably used the terminology of states and observations. For
        some environments such as the full-information games Chess and Go, these
        notions are equivalent. However, many environments prevent access to
        state information. For example, in many video games, information about
        the state is typically limited to audio-visual data, as underlying
        parameters are hidden from the agent. Likewise, quantum systems
        fundamentally limit accessible information about a quantum state. To
        distinguish this facet of MDPs, we define the following:
        \begin{definition*}\label{def:pomdp}
            A \textbf{partially observable Markov decision process} (POMDP)
            is the tuple\\ \((\mathcal{S},\ \mathcal{A},\ \mathcal{O},\
            \mathcal{T},\ \mathcal{R},\ \mathcal{B},\ \gamma)\) where:
            \begin{itemize}
                \item \(\mathcal{S}\) is a set of states
                \item \(\mathcal{A}\) is a set of actions
                \item \(\mathcal{O}\) is a set of observations
                \item {%
                    \(
                        \mathcal{T}:
                        \mathcal{S} \times \mathcal{S} \times \mathcal{A}
                        \longrightarrow [0, 1]
                    \)
                    is the state-transition probability:\\
                    \(
                        T^{a}_{ss'} = {
                            \mathds{P}\left[
                                s_{t + 1} = s' | s_{t} = s, a_{t} = a
                            \right]
                        }
                    \)
                }
                \item {%
                    \(
                        \mathcal{R}:
                        \mathcal{S} \times \mathcal{A}
                        \longrightarrow \mathds{R}
                    \)
                    is a reward function:\\
                    \(
                        R^{a}_{s} = {%
                            \mathds{E}[r_{t + 1}|s_{t} = s, a_{t} = a]
                        }
                    \)
                }
                \item {%
                    \(\mathcal{M}: \mathcal{S} \longrightarrow \mathcal{O}\)
                    is the observation function:\\
                    \(
                        \mathcal{M}_{so} = {%
                            \mathds{P}\left[o_{t} = o | s_{t} = s\right]
                        }
                    \)
                }
                \item \(\gamma \in [0, 1]\) is the discount factor
            \end{itemize}
            A POMDP has the Markovian property of independence from past and
            future states.
        \end{definition*}
        The (stochastic) observation function adds uncertainty to an agents
        decision-making, as it now needs to additionally learn a model
        distribution over the states:
        \begin{definition*}
            The \textbf{belief state} is a probability distribution over the
            states given a \textbf{history} \(h_{t} = \left(o_{0:t},
            a_{0:t-1}\right)\):
            \begin{equation*}
                b_{t}(s_{t}) = \mathds{P}\left[s_{t} = s | h_{t} \right]
            \end{equation*}
            The mapping from a history to the corresponding belief is given by a
            function:
            \begin{equation*}
                f^{*}(h_{t}) = b_{t}
            \end{equation*}
        \end{definition*}
        Such a belief is a sufficient statistic of the history (a compact
        summary of the history). It's update function can be calculated
        recursively from an initial observation:
        \begin{equation*}
            b'(s') = {%
                \frac{%
                    \mathcal{M}_{s'o'}
                    \underset{s\in \mathcal{S}}{\sum}\mathcal{T}^{a}_{ss'}b(s)
                }
                {Z}
            }
        \end{equation*}
        Here, \(Z\) is the normalisation factor. We obtain an update equation:
        \begin{equation*}
            b' = f(b\ |\ a, o')
        \end{equation*}
    \section{Reinforcement Learning Hyperparameters}\label{ssec:rl_hp}
        The learning rate, discount parameter (\cref{def:mdp}), and mini-batch
        size were chosen similar to Haarnoja~\textit{et
        al.}~\cite{haarnoja2019soft} and were tested and confirmed to be
        performative through a series of ablations. The experience replay
        capacity (i.e. the number of stored experiences) was reduced from 1M to
        100K as this smaller buffer did not alter performance. Likewise, the
        number of outputs and top outputs for the Q-networks is matched
        with~\cite{kuznetsov2020controlling}, and the dropout is matched
        with~\cite{hiraoka2022dropoutqfunctionsdoublyefficient}. These dropout
        and normalising layers help with agent stability and efficiency in
        training. A maximal and minimal log-variance is enforced to ensure
        training stability while also preventing convergence to a Dirac delta
        distribution. The Polyak parameter, \(\rho\), prevents large changes in
        the critics distribution and is chosen to be sufficiently small to
        stabilise training.
        \begin{table}[h]
            \centering
            \begin{tabular}{|c|c|c|}  
                \hline  
                \textbf{RL Parameter} & \textbf{Symbol} & \textbf{Value} \\
                \hline  
                Discount parameter & \(\gamma \) & 0.99 \\
                Polyak averaging parameter & \(\rho \) & 0.005 \\
                Number of outputs for Q-networks & \(q_{k}\) & 46 \\
                Number of top outputs used for TQC & \(q_{n}\) & 25 \\
                Learning rate & \(\eta \) & 5e-4 \\
                Dropout probability & p & 0.01 \\
                Mini-batch size & \(-\) & 256 \\
                Maximal log variance for policy & \(\log
                \nu_{\text{min}}\) & -15 \\
                Maximal log variance for policy & \(\log
                \nu_{\text{max}}\) & 4 \\
                Hidden layers & \(-\) & [512, 512] \\
                Relay capacity & \(-\) & 100,000 \\
                \hline  
            \end{tabular}
            \caption{%
                The parameters used for the reinforcement learning algorithm.
            }\label{tab:rl_hp}
        \end{table}
    \section{Unitary tomography}\label{ssec:u_t}
        Quantum process tomography (QPT) is a method of estimating and
        reconstructing a process (e.g. gates) from a collection of measurements
        on a quantum system. Process tomography poses a greater challenge than
        quantum state tomography (QST) (i.e. estimating a state), as multiple
        inputs are needed to fully and uniquely characterise a process. A
        general map has \(d^{2}(d^{2} - 1)\) independent parameters, as it
        requires \(d^{2}\) probe states to characterise a map
        \(\rho^{\text{out}}_{n} = \mathcal{P}(\rho^{\text{in}}_{n})\), which has
        \(d^{2} - 1\) independent parameters.

        Unitary gate tomography exploits the symmetries of \(U(n)\) to estimate
        \(d^{2} + d\) parameters of a unitary gate. This is achieved by
        utilising a specific set of informationally complete POVMs and input
        elements~\cite{baldwin2014quantum}. Flammia~\textit{et
        al.}~\cite{flammia2005minimal} show that the following set of \(2d\)
        POVMs is informationally complete for pure states:
        \begin{equation}\label{eq:povm}
            \begin{aligned}
                & E_{0} = a\vert 0\rangle\langle 0\vert \\
                & E_{j} = b\left(
                    1
                    + \vert j\rangle\langle 0\vert
                    + \vert 0\rangle\langle j\vert
                \right)\ \ \ \ j = 1, \ldots, d - 1\\
                & \widetilde{E}_{j} = b\left(
                    1
                    + i\vert j\rangle\langle 0\vert
                    - i\vert 0\rangle\langle j\vert
                \right)\ \ \ \ j = 1, \ldots, d - 1\\
                & E_{2d} = {%
                    1 - E_{0} - \sum_{j = 1}^{d - 1} E_{j} + \widetilde{E}_{j}
                }
            \end{aligned}
        \end{equation}
        Here, \(a, b > 0\) are chosen such that \(E_{2d}>0\). The input states
        must be chosen to account for the relative phases of the unitary
        elements. A suitable choice is~\cite{baldwin2014quantum}:
        \begin{equation}
            \begin{aligned}
                & \vert\psi_{0}\rangle = \vert 0\rangle \\
                & \vert\psi_{n}\rangle = \frac{1}{\sqrt{2}}\left(
                    \vert 0\rangle + \vert n\rangle
                \right)\ \ \ \ n = 1, \ldots, d - 1
            \end{aligned}
        \end{equation}
        The procedure for unitary tomography is as follows: Let U act on
        \(\ket{\psi_{0}}\) and perform QST on \(\ket{u_{0}} = U\ket{\psi_{0}}\)
        to characterise it. This involves finding the coefficients of the state,
        i.e. \(\ket{u_{0}} = \sum_{k}c_{0, k}\ket{k}\). Repeat for
        \(\ket{u_{1}}\) and use the relation
        \(U\ket{\psi_{1}}\bra{\psi_{1}}U^{\dagger}\ket{u_{0}} =
        \frac{1}{2}\left(\ket{u_{0}} + \ket{u_{1}}\right)\) to characterise
        \(\ket{u_{1}}\), including the relative phase to \(\ket{u_{0}}\). Repeat
        for all the inputs.

        To simulate a tomography experiment, we first extract the probabilities
        from an obtained unitary:
        \begin{equation}\label{eq:povm_probs}
            p_{n, j} = \text{Tr}\left(U\rho_{n}U^{\dagger}E_{j}\right)
        \end{equation}
        As an example, we have:
        \begin{equation}
            \begin{split}
                p_{0, 0} &= \bra{u_{0}}E_{0}\ket{u_{0}} = a|c_{0, 0}|^{2}\\
                p_{0, j} &= \bra{u_{0}}E_{j}\ket{u_{0}} = b\left(
                    1 + 2c_{0, 0}\text{Re}\left(c_{0, j}\right)
                \right)\\
                \widetilde{p}_{0, j} &= \bra{u_{0}}\widetilde{E}_{j}\ket{u_{0}}
                = b\left(1 + 2c_{0, 0}\text{Im}\left(c_{0, j}\right)\right)
            \end{split}
        \end{equation}
        Note that this procedure fails for \(c_{0, 0} = 0\). Then, we perform a
        multinomial sampling of \(N_{m}\) shots along \(d - 1\) input axes. This
        number bounds the precision of the fidelity. Finally, we reconstruct the
        unitary from our obtained samples using an inversion
        of~\cref{eq:povm_probs} with an approximated \(p_{n, j}\). Additionally,
        we map the reconstruction to the closest unitary in order to account for
        sampling size effects, i.e. when \(N_{m}\) is small, the estimated gate
        is not unitary.

\begin{figure}
            \centering
            \includegraphics[scale=0.250]{%
                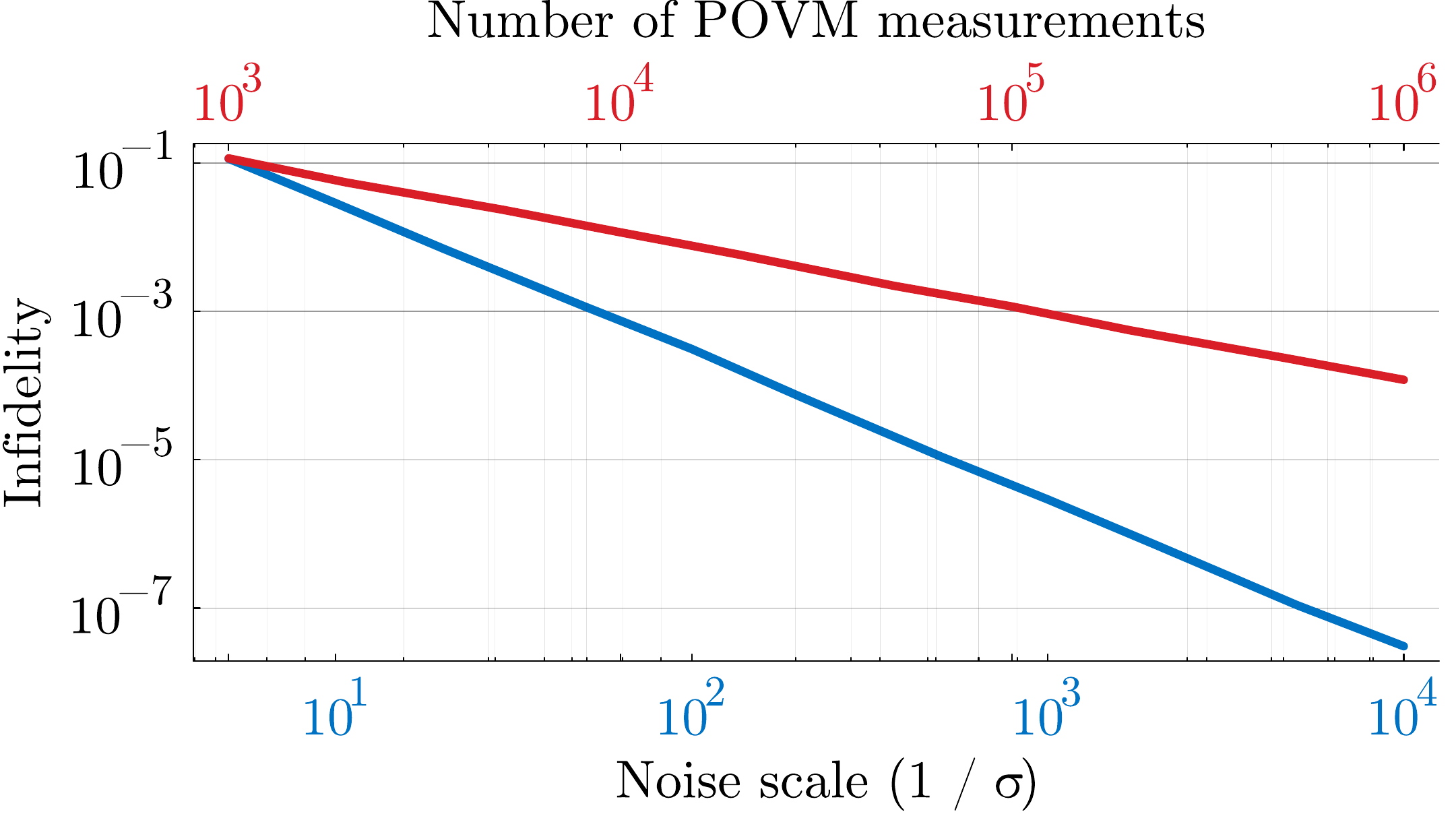
            }
            \caption{%
                The infidelity scaling of a random \(6d\) unitary using a
                POVM-based approach (red) and a simple additive Gaussian noise
                (blue). We use a linear fit to heuristically approximate the
                number of shots to the (inverse) standard deviation that
                achieves the same infidelity.
            }\label{fig:gauss_noise}
        \end{figure}

        Due to the exponential scaling of the infidelity between the actual gate
        and its estimated reconstruction with respect to the number of POVM
        measurements (\cref{fig:gauss_noise}), it is challenging to reach a
        desired infidelity threshold. For example, \(10^{5}\) measurements are
        required to reach an infidelity of \(10^{-3}\), hence is both
        computationally expensive if simulated and experimentally costly if
        performed. As a result, we devise an alternative approach that is less
        computationally costly.

        We add element-wise Gaussian noise of a certain standard deviation to
        the actual gate to approximate the shot noise of a POVM-based
        estimation. As before, we map the noisy unitary to the closest unitary
        to account for sampling size effects. By calculating the mean infidelity
        for a given a standard deviation, we extrapolate an infidelity scaling.
        Combined with the POVM-based infidelity scaling, we map a standard
        deviation to the approximate number of measurements required to achieve
        the same infidelity (\cref{fig:gauss_noise}). Therefore, the
        computationally expensive limit of a large number of measurements is
        approximated with a cheaper additive Gaussian noise of low standard
        deviation.
    \section{Single qubit noise-robust phase gate}\label{ssec:1dq}
\begin{figure}[h]
            \centering
            \includegraphics[scale=1.0]{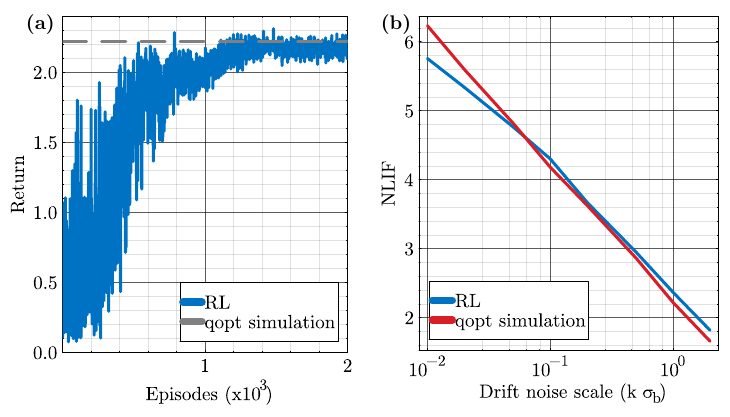}
            \caption{%
                (a) Final infidelity performance of both an RL agent and a qopt
                based simulation with respect to the drift noise scale on a
                simple 1-qubit system. The performance scaling indicates RL is
                only bounded by a fundamental infidelity scaling due to the
                noise and not any optimisation issues. (b) The maximum training
                curves for an RL agent in a 1-qubit environment with only drift
                noise and with a noise-averaged reward of negative log
                infidelities. The RL agent will perform similarly with "full"
                observation (i.e. exact tomography) to gradient-based approach
                (of a protocol length of 10ns and 20 inputs) using the qopt
                package.
            }
            \label{fig:qopt_rl_fidelity}
        \end{figure}

        We use a noise-averaged reward with exact tomography as
        in~\cref{ssec:robust_exact} for a simple one-qubit quantum dot system
        that only consists of hyperfine noise. Similar to the more complicated
        two-qubit case, we see in~\cref{fig:qopt_rl_fidelity} (a) that we obtain
        a final fidelity less than the desired 99.9\%, with an achieved
        log-infidelity of about 2.1. However, like the two-qubits, our
        gradient-based qopt simulation has a similar performance, additionally
        indicating a fidelity limit that is bounded by the noise scale and not
        on agent limitations, as can be seen in~\cref{fig:qopt_rl_fidelity} (b),
        where train an RL agent and perform a qopt simulation at different noise
        scales.

    \pagebreak


\end{document}